\documentclass[aps,prb,reprint,groupedaddress,showpacs]{revtex4-1}
\usepackage[dvipdfm]{graphicx}
\usepackage{amsmath}
\usepackage{txfonts}
\usepackage{bm}

\begin{document}

%\preprint{}

\title{Skyrme Crystal in bilayer and multilayer graphene}


\author{Yasuhisa Sakurai}
%\thanks{}
\affiliation{Department of Basic Science, The University of Tokyo, 3-8-1 Komaba, Tokyo 153-8902, Japan}
\author{Daijiro Yoshioka}
\affiliation{Department of Basic Science, The University of Tokyo, 3-8-1 Komaba, Tokyo 153-8902, Japan}


\date{\today}

\begin{abstract}
The ground state of the two-dimensional electron systems in Bernal bilayer and ABC-stacked multilayer graphenes in the presence of a strong magnetic field is investigated with the Hartree-Fock approximation. 
Phase diagrams of the systems are obtained, focusing on charge density wave states including states with vortices of valley pseudospins (called a Skyrme crystal). 
The single-electron states in these stacked graphenes are given by two-component wave functions. 
That of the first excited Landau level has the same component as the lowest Landau level of the ordinary two-dimensional electrons. 
Because of this localized wave functions, 
the Skyrme crystal has low energy in this first excited level up to four layers of graphene, 
when the inter-layer distance is assumed to be infinitesimal. 
At the same time, bubble crystals are suppressed, so the phase diagram is different from that of a single-layer graphene.

\end{abstract}

%\keywords{}
\maketitle
\section{introduction}
Over the past decades, the conventional two-dimensional electron system (2DES) in semiconductor heterostructures in a strong magnetic field has been studied. 
It is found that electron-electron interaction brings various phases [e.g., fractional quantum Hall effect (FQHE) states~\cite{PhysRevLett.48.1559, PhysRevLett.50.1395} 
and charge density wave (CDW) states~\cite{PhysRevB.19.5211}]. 
Similarly, graphene,~\cite{RevModPhys.81.109} 
a flat sheet of carbons with a honeycomb lattice exfoliated from a graphite in 2004,~\cite{Sci.306.666} is under intensive investigation as a new 2DES whose electron has a valley degree of freedom ($K, K'$) and a linear dispersion. 
Exhibition of FQHE~\cite{PhysRevLett.97.126801, JPSJ.78.104708} and CDW~\cite{PhysRevB.75.245414, PhysRevB.78.085309} states in the new 2DES has been expected; FQHE was recently observed in a single-layer graphene (SLG).~\cite{nat462.192,nat462.196} 
Furthermore, few-stacked graphenes have recently attracted attention due to their intriguing properties: a band structure tunable by the number of layers and their stacking sequence, and a band gap controllable by a perpendicular electrical field.~\cite{PhysRevB.77.155416, ProgTheorPhysSuppl.176.227} 

We focus on Bernal bilayer and ABC-stacked multilayer graphenes, 
which have chiral electrons concentrated on outer layers in low-energy states. 
In this paper, the CDW ground state of the 2DES in bilayer graphene (BLG) and multilayer graphene ($M$-LG, $M \ge 3$) in a strong magnetic field is studied with the Hartree-Fock approximation. 
As candidates of the ground state, 
the electron (hole) Wigner crystal, electron (hole) bubble crystal, and Skyrme crystal are considered. 
A Skyrme crystal is the state with a topological texture of valley pseudospins. 
It has the lowest energy around a filling $\nu=1$ in SLG.~\cite{PhysRevB.78.085309} 
Our mean-field analysis predicts that the valley Skyrme crystal also has low energy in BLG and $M$-LG ($M = 2, 3, 4$) at the first excited Landau level when the interlayer distance $d$ is assumed to be infinitesimal. 

This paper is organized as follows. 
In the next section, we set up a low-energy effective Hamiltonian in the presence of a  magnetic field for SLG, BLG, and $M$-LG. 
Then, the Hartree-Fock approximation is applied to the system of interacting electrons and  self-consistent equations are derived. 
In Sec. \ref{sec:CDW}, CDW states are introduced, including Skyrme crystals (more properly, the meron and meron pair crystal). 
As a preliminary calculation, the energy of a skyrmion (antiskyrmion) pair excitation at a $\nu=1$ ferromagnetic state is evaluated and compared with that of a separated particle-hole excitation to find the condition where the Skyrme crystal has low energy.
In Sec. \ref{sec:Results}, numerical results are presented at the Landau level where the  Skyrme crystal is expected. 
In Sec. \ref{sec:Discussion}, the validity of the results and a relation to experiments are discussed. 

\section{Model\label{sec:Model}}
\subsection{Single-particle state in a magnetic field}
For a single-layer graphene (SLG), 
the low-energy effective Hamiltonian around the $K$ valley is given by
\begin{equation}
\mathcal{H}_{K}^{\textit{SLG}} = v_F ( p_x \sigma_x + p_y \sigma_y ),
\label{eq:SLGHamiltonian}
\end{equation}
which has an electron with a linear dispersion.~\cite{RevModPhys.81.109,PhysRev.71.622, PhysRev.109.272,PhysRevLett.53.2449,PhysRevLett.61.2015, JPSJ.74.777, PhysRevB.73.125411} 
Here, $v_F$ is Fermi velocity, 
and $\sigma_x$ and $\sigma_y$ are Pauli matrices acting on sublattice (A, B) space. 
The Hamiltonian for the $K'$ valley is $\mathcal{H}_{K'}^{\textit{SLG}} = - ( \mathcal{H}_{K}^{\textit{SLG}} )^*$. 
In a perpendicular magnetic field $\bm{B} = B \bm{e}_z$, 
using a magnetic ladder operator $a = (\pi_x - i \pi_y ) l_B / \sqrt{2} \hbar$
($\bm{\pi} = \bm{p} - e \bm{A}$), 
the single-particle Hamiltonian is written as
\begin{equation}
\mathcal{H}_{K}^{\textit{SLG}} =  \frac{\sqrt{2} \hbar v_F}{l_B}
	\begin{pmatrix}
	0 & a \\
	a^{\dagger} & 0 \\
	\end{pmatrix},
\end{equation}
where $l_B = \sqrt{\hbar/eB}$ is magnetic length. 

The eigenenergies are 
\begin{equation}
E_N = \pm v_F \sqrt{2 e \hbar  B N},\ \ ( N=0,1,\dots),
\end{equation}
where a positive (negative) sign is taken as the electron (hole) state. 
The corresponding eigenstates have the form
\begin{equation}
\Phi_{K,N,X} ^{\textit{SLG}}(\bm{r}) = \begin{cases}
 \frac{1}{\sqrt{2}} e^{-i \bm{K} \cdot \bm{r} }\begin{pmatrix} 
                   \phi_{N-1,X}(\bm{r}) \\ 
                   \pm \phi_{N,X} (\bm{r}) \\ 
                    \end{pmatrix}  &\ \   (N \ge 1), \\
\ \ \   e^{-i \bm{K} \cdot \bm{r} } \begin{pmatrix} 
                   0 \\ 
                    \phi_{N,X} (\bm{r}) \\ 
                    \end{pmatrix}  &\ \  (N = 0), \\
\end{cases}
\end{equation}
for the $K$ valley. 
In the Landau gauge, $\phi_{N,X}(\bm{r})$ is given by
\begin{equation}
\phi_{N,X}(\bm{r}) = \left( \frac{1}{\sqrt{\pi} 2^N N! l_B L} \right)^{1/2} i^Ne^{ -i \frac{X}{l_B^2} y - \frac{(x-X)^2}{2l_B^2}} H_N \left( \frac{x-X}{l_B} \right), 
\label{eq:singlewf}
\end{equation}
where $L$ is the length of the system, $H_N(x)$ is the Hermite polynomial, 
and $X= k_y l_B^2$ is the guiding center coordinate. 
A macroscopic number of states with different $X$ are degenerated at one Landau level. 
A single-particle density on each sublattice $2 \pi l_B^2\phi_{N, X}^* \phi_{N X}$ becomes broader as index $N$ increases. 
The states with broad density induce a bubble CDW and lose the profit to form skyrmions (see Sec. \ref{sec:CDW}). 

Bernal-stacked bilayer graphene (BLG) and ABC-stacked multilayer graphene~\cite{note1} ($M$-LG) have low-energy quasiparticles with a dispersion $E \propto p^M$, 
where $M$ ( $\ge 2$) is the number of layers.~\cite{PhysRevB.77.155416,ProgTheorPhysSuppl.176.227,
PhysRevB.73.214418, PhysRevLett.96.086805} 
In two adjacent layers, 
BLG and $M$-LG have interlayer hoppings $t_\perp$ between the sublattice $\beta$ in the lower layer and the sublattice $\alpha$ in the upper layer.
In a magnetic field, 
the effective single-particle Hamiltonian for BLG and $M$-LG, 
which act on the outermost layer (top, bottom) space, have the same form 
\begin{equation}
\mathcal{H}_{K}^{\textit{$M$-LG}} =  \hbar \omega_M
	\begin{pmatrix}
	0 & a^M \\
	(a^{\dagger})^M & 0 \\
	\end{pmatrix}, \ \ (  M \ge 2),
\end{equation}
where off-diagonal components represent $M$-times hopping via high-energy states in  the inner layers. 
Its eigenenergy is
\begin{equation}
E_N = \pm \hbar \omega_M \sqrt{N (N-1) \cdots (N-M+1)},
\end{equation}
where $\hbar \omega_M = t_{\perp} ( \sqrt{2} \hbar v_F / t_{\perp} l_B )^M \propto  B^{M/2}$.
At zero energy, $M$ Landau levels are degenerated.  
The corresponding eigenstate has the form
\begin{equation}
\Phi_{K,N,X}^{M\textit{-LG}} (\bm{r}) = \begin{cases}
 \frac{1}{\sqrt{2}}  e^{-i \bm{K} \cdot \bm{r} }  \begin{pmatrix} 
                   \phi_{N-M,X}(\bm{r}) \\ 
                   \pm \phi_{N,X} (\bm{r}) \\ 
                    \end{pmatrix}  &\ \   (N \ge M), \\
\ \ \  e^{-i \bm{K} \cdot \bm{r} } \begin{pmatrix} 
                   0 \\ 
                    \phi_{N,X} (\bm{r}) \\ 
                    \end{pmatrix}  &\ \  (N<M), \\
\end{cases}\label{eq:MLGwf}
\end{equation}
for the $K$ valley. 
Notice that in the first excited state realized at $N=M$, the upper 
component of the wave function is given by $\phi_{0,X}(\bm{r})$, which 
is the ground-state wave function of the ordinary 2$d$ electrons. 

At Landau level $N < M$, the valley degree of freedom coincides with the layer degrees of freedom. 
If the layer degree of freedom in the $M$-LG model is regarded as the sublattice degree of freedom, 
Eq. (\ref{eq:MLGwf}) is formally identical to that of the SLG model when $M=1$. 
In the following,  the case of  $M=1$ ($M=2$) represents the model of SLG (BLG).

\subsection{Hartree-Fock Hamiltonian}
Although the electron state is mainly considered in the following, the hole state can be treated in the same way. 
In the present case of strong magnetic fields, 
electronic spins are completely polarized 
and the gap around the neutrality point is sufficiently large due to exchange enhancement; 
thus we take only one spin component and ignore the effect of Landau-level transitions. 

The interaction between the electrons in different layers is weakened by an interlayer distance $d$ ($\sim3.35$ \AA \  for BLG). 
When $B = 10$ T, $d$ is quite small compared to magnetic length $l_B \sim 80 $ \AA,    
so we consider the vanishing limit of interlayer distance as an approximate model. 

Apart from a constant kinetic term, 
the Hamiltonian for interacting electrons in $M$-LG ($M \ge 1$) is given by 
\begin{equation}\begin{split}
\mathcal{H}_{\text{int} } = \frac{1}{2}& \int d^2 \bm{r}
 \int d^2 \bm{r}' \sum_{\sigma_1,\sigma_2,\sigma_3,\sigma_4} 
  \hat{\psi}_{M,N}^{\sigma_1 \dagger}(\bm{r}) \hat{\psi}_{M,N}^{\sigma_2 \dagger}(\bm{r}') \\
& \times V(|\bm{r}-\bm{r}'|) \hat{\psi}_{M,N}^{\sigma_3}(\bm{r}') \hat{\psi}_{M,N}^{\sigma_4}(\bm{r}), 
\label{HFHamiltonian1}
\end{split}\end{equation}
where the field operator $\hat{\psi}_{M,N}^{\sigma}(\bm{r})$ is represented by
\begin{equation}
\hat{\psi}_{M,N}^{\sigma}(\bm{r}) 
= \sum_X \Phi_{\sigma, N, X}^{M\text{-LG}} (\bm{r}) \hat{c}_{N,X}^{\sigma}, 
\end{equation}
and Coulomb interaction is written as 
\begin{equation}
V ( r) =  \frac{e^2} { \epsilon r}. 
\end{equation}
Here, $\epsilon$ is the dielectric constant, 
$\sigma = \pm $ represents valley $K$ and $K'$, respectively, 
$\hat{c}_{N,X}^{\sigma}$ is an annihilation operator of the electron at valley $\sigma$ with  guiding center $X$. 
A valley scattering term is relatively small when the cutoff $q_{max} \ll K$ is used,~\cite{PhysRevB.75.245414, PhysRevB.78.085309, PhysRevB.74.161407} 
since $l_B \gg  a$ ( $a = 1.42 $ \AA \  is the lattice constant), 
so we can ignore it as far as the CDW with magnetic length scale is concerned. 
Then, the Hamiltonian in $\bm{q}$ space is written as
\begin{equation}
\mathcal{H}_{\text{int} } 
= \frac{1}{2 L^2} \sum_{\bm{q}} 
\sum_{\sigma_1,\sigma_2}
 \sum_{\xi_1, \xi_2 } V(q) \rho^{\sigma_1}_{\xi_1} (-\bm{q} )
 \rho^{\sigma_2}_{\xi_2} (\bm{q} ) ,
\end{equation}
\begin{equation}\begin{split}
&\rho^{\sigma}_{\xi} (\bm{q} ) \\&=   \frac{1}{2} \sum_X e^{ -iq_x X - \frac{1}{4} q^2 l_B^2 }  L_{N_{\sigma, \xi}}\left( q^2 l_B^2 / 2 \right) \hat{c}_{N,X+\frac{1}{2}q_y l_B^2}^{\sigma \dagger}  \hat{c}_{N,X-\frac{1}{2}q_y l_B^2}^{\sigma}, 
\label{eq:density} \end{split}\end{equation}
\begin{equation}
 V (q) = \frac{2 \pi e^2}{\epsilon q}, 
\end{equation}
where $L_n(x)$ is the Laguerre polynomial, 
and $N_{\sigma, \xi}$ is defined by  
$N_{+, \uparrow} = N_{-, \downarrow} = N-M$ and $N_{+, \downarrow} = N_{-, \uparrow} = N$. 
When Landau-level index $N$ is smaller than the number of layers $M$, 
the valley degree of freedom coincides with that of the layers (sublattices for SLG), 
and $\rho^{+}_{\uparrow}(\bm{r}) = \rho^{-}_{\downarrow}(\bm{r}) =0 $ and  
$\rho^{+}_{\downarrow}(\bm{r})$ and $\rho^{-}_{\uparrow}(\bm{r})$ are twice the amount of Eq. (\ref{eq:density}).

For the Hartree-Fock decoupling of $\mathcal{H}_{\text{int}}$, 
we assume the following order parameters: 
\begin{equation}
\Delta^{\sigma, \sigma'}_N(\bm{Q}) = \frac{2\pi l_B^2}{L^2} \sum_X 
\langle \hat{c}_{N, X_+}^{\sigma \dagger}\hat{c}_{N, X_-}^{\sigma'} \rangle
e^{-iQ_x X},
\end{equation}
where $\bm{Q}$ is a reciprocal lattice vector of the CDW state. 
Then the Hartree-Fock Hamiltonian is given by~\cite{PhysRevLett.65.2662,PhysRevB.44.8759,PhysRevB.46.10239,PhysRevB.45.11054}
\begin{equation}\begin{split}
\mathcal{H}_{\text{HF}} = &  \sum_{\bm{Q}} \sum_X \sum_{\sigma \sigma'}  H_{M,N}^{\sigma, \sigma'}(Q)e^{-iQ_xX} \Delta_N^{\sigma, \sigma} (-\bm{Q} )\hat{c}_{N, X_+}^{\sigma' \dagger}\hat{c}_{N, X_-}^{\sigma'} \\
& -\sum_{\bm{Q}} \sum_X \sum_{\sigma \sigma'} X_{M,N}^{\sigma, \sigma'} (Q) e^{-iQ_xX}\Delta_N^{\sigma, \sigma'} (-\bm{Q}) \hat{c}_{N, X_+}^{\sigma' \dagger}\hat{c}_{N, X_-}^{\sigma} ,  
\label{HFH}
\end{split}\end{equation}
where $X_{\pm} =X \pm Q_y l_B^2 /2$.
The Hartree-Fock potential consists of a direct term 
\begin{equation}
H_{M,N}^{\sigma, \sigma'}(Q)=\sum_{\xi,\xi'} H_{M,N}^{\sigma \xi, \sigma' \xi'}(Q),
\end{equation} 
\begin{equation}\begin{split}
&H_{M,N}^{\sigma \xi, \sigma' \xi'}(Q)  \\& 
=   \frac{ 1 }{4} \frac{e^2}{ \epsilon l_B}  \frac{1}{Q l_B}  L_{N_{\sigma \xi}} \left( Q^2l_B^2/2 \right) L_{N_{\sigma' \xi'}} \left( Q^2l_B^2/2 \right)  e^{ - Q^2 l^2 /2 },
\end{split}\end{equation}
and an exchange term
\begin{equation}
X_{M,N}^{\sigma, \sigma'}(Q)=\sum_{\xi,\xi'} X_{M,N}^{\sigma \xi, \sigma' \xi'}(Q), 
\end{equation}
\begin{equation}\begin{split}
X_{M,N}^{\sigma \xi , \sigma' \xi'}(Q) 
 =  & \frac{1}{4 }  \frac{e^2}{\epsilon l_B}
 \int_0^{\infty} d x 
 J_0 (Q x l_B)  e^{-\frac{1}{2} x^2 } \\
& \times L_{N_{\sigma \xi }} \left( \frac{x^2}{2}\right)
L_{N_{\sigma' \xi' }}  \left( \frac{x^2}{2}\right) .
\end{split}\end{equation}
Here, $J_n(x)$ is a Bessel function of the first kind. 

The real-space density of electrons at valley $\sigma$ and layer (sublattice) $\xi$ is given by
\begin{equation}
\rho^{\sigma}_{\xi}(\bm{r}) =\frac{1}{2}
\frac{1}{2 \pi l_B^2} \sum_{ \bm{Q}}
\Delta_N^{\sigma,\sigma} (\bm{Q}) L_{N_{\sigma,\xi }}(Q^2 l_B^2/2)
e^{i \bm{Q} \cdot \bm{r} - Q^2l^2/4}. 
\end{equation}
The filling factor at Landau level $N$ is defined as
\begin{equation}
\nu_N = \frac{N_e}{N_\phi} \in [0, 2], 
\end{equation}
where $N_\phi = S/ 2 \pi l_B^2$ ($S$ is a area of the system) is the degeneracy of Landau orbitals and $N_e$ is the number of electrons. The factor 2 comes from the valley degree of freedom.

\subsection{Green's function method}
To determine the order parameters self-consistently, 
the Green's function method is used.~\cite{PhysRevB.75.245414,PhysRevB.78.085309,PhysRevB.44.8759,PhysRevB.46.10239} 
The single-particle Matsubara Green's function is defined by
\begin{equation}
G_N^{\sigma , \sigma '} (X, X',\tau ) = - \langle T c^{\sigma}_{N,X} (\tau) c^{\sigma' \dagger}_{N,X'} (0) \rangle .
\end{equation}
The Fourier transformation
\begin{equation}\begin{split}
&G_N^{\sigma,\sigma'} (\bm{Q}, \tau ) \\ &= \frac{2 \pi l_B^2}{ L^2} \sum_{X,X'}
  e^{-(i/2)Q_x(X+X') } \delta_{X,X'-Q_yl^2} G_N^{\sigma, \sigma'}(X,X',\tau), 
\end{split}\end{equation}
relates to the order parameters $\Delta_N^{\sigma' \sigma} (\bm{Q})$ by
\begin{equation}
\Delta_N ^{\sigma' \sigma} (\bm{Q}) = G_N^{\sigma,\sigma'} (\bm{Q},\tau = 0^-) .
\end{equation}
The equation of motion for $G_N^{\sigma,\sigma'} (\bm{Q}, \tau )$ is derived as
\begin{equation}\begin{split}
&\hbar ( i \omega_n - \mu ) G_N^{\sigma,\sigma'} (\bm{Q},\omega_n) - \hbar \delta_{\bm{Q},0} \delta_{\sigma,\sigma'} \\
& = \sum_{\sigma''} \sum_{\bm{Q}'}
 \Sigma_{M, N}^{\sigma,\sigma''}(\bm{Q},\bm{Q}')  G_N^{\sigma'',\sigma'} (\bm{Q}',\omega_n) , 
\end{split}\label{GFeom}
\end{equation}
from the Heisenberg equation. Here, $\omega_n$ is the Matsubara frequency for fermions, 
and the self-energy $\Sigma_{M, N}^{\sigma,\sigma'}(\bm{Q},\bm{Q}') $ is given by
\begin{equation}\begin{split}
&\Sigma_{M, N}^{\sigma,\sigma'}(\bm{Q},\bm{Q}')  \\& =  \sum_{\sigma''} H_{M, N}^{\sigma'' \sigma}(\lvert \bm{Q}-\bm{Q}' \rvert ) e^{-\bm{Q} \times \bm{Q}' l^2/2}
\Delta_N^{\sigma'',\sigma''}(\bm{Q}-\bm{Q}') \delta_{\sigma,\sigma'} \\
& \ \ \ \ -X_{M, N}^{\sigma' \sigma} (\lvert \bm{Q}-\bm{Q}' \rvert ) e^{-\bm{Q} \times \bm{Q}' l^2/2}
 \Delta_N^{\sigma',\sigma}(\bm{Q}-\bm{Q}'). 
\end{split}\label{selfenergycomp}\end{equation}
To solve this self-consistent equation, 
we diagonalize the self-energy matrix 
\begin{equation}\begin{split}
&\sum_{\bm{Q}'} 
	\begin{pmatrix}
	\Sigma_{M, N}^{+,+}(\bm{Q},\bm{Q}') & \Sigma_{M, N}^{+,-}(\bm{Q},\bm{Q}') \\
	\Sigma_{M, N}^{-,+}(\bm{Q},\bm{Q}') & \Sigma_{M, N}^{-,-}(\bm{Q},\bm{Q}') \\
	\end{pmatrix}
	\begin{pmatrix}
	V_j^+(\bm{Q}') \\
	V_j^-(\bm{Q}') \\
	\end{pmatrix} \\
&\ \  = \gamma_j
	\begin{pmatrix}
	V_j^+(\bm{Q}) \\
	V_j^-(\bm{Q}) \\
	\end{pmatrix}, 
\label{eq:eigeneq}
\end{split}\end{equation}
where $(V_j^+,V_j^-)$ is the $j$th eigenvector with eigenvalue $\gamma_j$.
The order parameters are obtained from the eigenvectors and eigenvalues
\begin{equation}
\Delta^{\sigma \sigma'}_N (\bm{Q}) 
= \sum_k f ( \gamma_{k} - \mu )  V_{k}^{\sigma'} (\bm{Q})  V_{k}^{\sigma *}(0), 
\label{eq:odprmt}
\end{equation}
where $f(x)$ is the Fermi-Dirac distribution function. 
The chemical potential $\mu$ is determined from
\begin{equation}
\sum_{\sigma} \Delta^{\sigma \sigma}_N (0) = \sum_{\sigma, j} V_j^\sigma(0) V_j^{\sigma *}(0) f(\gamma_j -\mu) = \nu_N .
\end{equation}
Self-consistent equations are numerically calculated to yield the order parameters and the Hartree-Fock energy per particle for several CDW states introduced in Sec. \ref{sec:CDW}. 

The order parameter sum rule at zero temperature,~\cite{PhysRevB.27.4986} 
extended to the case of valley degeneracy 
\begin{equation}
\begin{split}
\sum_{\bm{Q}} \sum_{\sigma'} \lvert \Delta^{\sigma \sigma'}_N (\bm{Q}) \rvert ^2 & = \Delta^{\sigma \sigma}_N (0) = \nu_N^\sigma, 
\end{split}
\end{equation}
is easily derived from Eq. (\ref{eq:odprmt}).
Here, $\nu_N^{\sigma}$ is the contribution from valley $\sigma$ electrons to the partial filling factor $\nu_N$. 
This relation is used to check convergence of the results.

\section{Charge Density Wave States\label{sec:CDW}}
\subsection{Valley skyrmion}
It is useful to map the valley degrees of freedom to a pseudospin.~\cite{PhysRevB.78.085309} 
In this language, the components of the pseudospin vector density 
$\bm{P}(\bm{Q}) = P_x ( \bm{Q} ) \hat{x} +P_y ( \bm{Q} ) \hat{y} +P_z ( \bm{Q} ) \hat{z}$ are defined by
\begin{equation}
P_x(\bm{Q}) = \frac{\Delta^{+-}_N (\bm{Q}) + \Delta^{-+}_N (\bm{Q}) }{2},
\end{equation}
\begin{equation}
P_y(\bm{Q}) = \frac{\Delta^{+-}_N (\bm{Q}) - \Delta^{-+}_N (\bm{Q}) }{2i},
\end{equation}
\begin{equation}
P_z(\bm{Q}) = \frac{\Delta^{++}_N (\bm{Q}) - \Delta^{--}_N (\bm{Q}) }{2}.
\end{equation}
In this paper, states with a topological pseudospin texture, which is called a skyrmion,~\cite{SandI} are considered. 
A skyrmion is a kind of spin texture usually used to describe a magnetic order and 
was first introduced in hadron physics.~\cite{NuclPhys.31.556} 
For conventional and SLG 2DES, 
a state with aligned skyrmions, 
which is called a Skyrme crystal, 
has been shown theoretically to be the ground state around $\nu=1$.~\cite{PhysRevB.78.085309, SurfSci.361} 

\subsection{Preliminary considerations\label{subsec:excitation}}
The excitation energy of a skyrmion in a ferromagnetic uniform state at $\nu_N =1$ is evaluated to find the conditions on which the Skyrme crystal is preferred.~\cite{PhysRevB.51.5138, PhysRevB.74.075423}
When the interlayer distance $d\to0$, 
the energy of a skyrmion (antiskyrmion) pair excitation $\Delta_{\textit{SK}}$ 
and the energy of a widely separated particle-hole pair excitation $\Delta_{\textit{PH}}$ are given by 
\begin{equation}
\Delta_{\textit{SK}} = \frac{1}{4 \pi} \int_0^{\infty} q^3 V(q)  \left[  \mathcal{F}_{M,N} (q) \right] ^2
e^{-q^2 l_B^2 /2} dq,
\end{equation}
\begin{equation}
\Delta_{\textit{PH}} = \frac{1}{2 \pi} \int_0^{\infty} q V(q)  \left[  \mathcal{F}_{M,N} (q) \right] ^2
e^{-q^2 l_B^2 /2} dq ,
\end{equation}
where the form factor $\mathcal{F}_{M,N} ( q )$ has the form
\begin{equation}
\mathcal{F}_{M,N} ( q ) =\begin{cases}
\frac{1}{2} \left[ L _{ N } \left( \frac{q^2l_B^2}{2} \right)+
L _{ N -M} \left( \frac{q^2l_B^2}{2} \right) \right] 
 & (N \ge M) \\
L_N \left( \frac{q^2l_B^2}{2} \right)  & (N < M) \end{cases}. 
\end{equation}
The ratio of the two energies of the pair excitations is the same as 
that of single-particle excitations, from particle-hole symmetry. 
For conventional, SLG, BLG, and $M$-LG ($M = 3, 4, 5$) 2DESs, 
these energies at Landau level $N \le 5$ are presented in Table~\ref{tableexcitation}. 
It shows that skyrmion excitation is favored (1) at $N=0$ in all systems as single-particle wave functions being identical to conventional one, (2) at $N=1,2,3$ in SLG, and (3) at $N= M$ in $M$-LG ($M = 2, 3, 4$). 
Thus we can expect that the Skyrme crystal becomes the ground state in these situations around $\nu_N=1$.

\begin{table}[t]
\caption{Hartree-Fock quasiparticle and skyrmion (antiskyrmion) particle-hole excitation gaps (in unit of $e^2/\epsilon l_B \sqrt{\pi / 2 }$) at Landau level $N$ 
for 2DES in a conventional semiconductor structure ($\Delta_{PH}^{conv.}$ and $\Delta_{SK}^{conv.}$), 
SLG ($\Delta_{PH}^{SLG}$ and $\Delta_{SK}^{SLG}$), 
BLG ($\Delta_{PH}^{BLG}$ and $\Delta_{SK}^{BLG}$), 
tri-LG ($\Delta_{PH}^{3-LG}$ and $\Delta_{SK}^{3-LG}$), 
tetra-LG ($\Delta_{PH}^{4-LG}$ and $\Delta_{SK}^{4-LG}$), 
and penta-LG ($\Delta_{PH}^{5-LG}$ and $\Delta_{SK}^{5-LG}$). 
For conventional and SLG 2DES, 
the energies of the two excitations were compared by Yang \textit{et al.}~\cite{PhysRevB.74.075423} 
The situation where the skyrmion is favored is emphasized by thick ``$\bm{>}$''. }
\label{tableexcitation}
\begin{center} 
\begin{tabular}{ c c c c c c c c c c}
\hline
 $N$ & $\Delta_{PH}^{conv.}$ & & $\Delta_{SK}^{conv.}$ & $\Delta_{PH}^{SLG}$ & & $\Delta_{SK}^{SLG}$&  $\Delta_{PH}^{BLG}$ & & $\Delta_{SK}^{BLG}$  \\
\hline
 0  & 1 & $\bm{>}$ & 1/2 & 1 & $\bm{>}$ & 1/2  &  1 & $\bm{>}$ & 1/2  \\
 1 & 0.75  & $<$ & 0.875  & 0.6875  & $\bm{>}$ &  0.2188 &  0.75  & $<$ & 0.875 \\
 2 & 0.6406  & $<$ & 1.1328  & 0.5664 & $\bm{>}$ & 0.3301 & 0.5977  & $\bm{>}$ & 0.3770 \\
 3 & 0.5742  & $<$ & 1.3418  & 0.5029 & $\bm{>}$ & 0.4097 & 0.5029  & $<$ & 0.5151\\
 4 & 0.5279  & $<$ & 1.5522  & 0.4608 & $<$ & 0.4754 & 0.4528  & $<$ & 0.6181\\
 5 & 0.4927  & $<$ & 1.6834  & 0.4298 & $<$ & 0.5328  & 0.4187  & $<$ & 0.7048 \\ \hline
\end{tabular}%

\begin{tabular}{ c c c c c c c c c c}
\hline
 $N$ & $\Delta_{PH}^{3-LG}$ & & $\Delta_{SK}^{3-LG}$ & $\Delta_{PH}^{4-LG}$ & & $\Delta_{SK}^{4-LG}$&  $\Delta_{PH}^{5-LG}$ & & $\Delta_{SK}^{5-LG}$  \\
\hline
 0  & 1 & $\bm{>}$ & 1/2 & 1 & $\bm{>}$ & 1/2  &  1 & $\bm{>}$ & 1/2  \\
 1 & 0.75  & $<$ & 0.875  & 0.75  & $<$ &  0.875 &  0.75  & $<$ & 0.875 \\
 2 & 0.6406  & $<$ & 1.1328  & 0.6406 & $<$ & 1.1328 & 0.6406  & $<$ & 1.1328 \\
 3 & 0.5498  & $\bm{>}$ & 0.4448  & 0.5742 & $<$ & 1.3418 & 0.5742  & $<$ & 1.3418\\
 4 & 0.4660  & $<$ & 0.5808  & 0.5285 & $\bm{>}$ & 0.4870 & 0.5279  & $<$ & 1.5522\\
 5 & 0.4223  & $<$ & 0.6829  & 0.4406 & $<$ & 0.6284  & 0.4963  & $<$ & 0.5390 \\ \hline
\end{tabular}%
\end{center}
\end{table}

The reason why a skyrmion can be a low-energy excitation is the following.~\cite{QHE}  
The ground state at $\nu_N =1$ is a pseudospin ferromagnetic liquid state. 
When a hole is introduced in this state without flipping pseudospin 
of other electrons, the charge density of the hole is given by an 
eigenstate of the angular momentum, and is concentrated. On the other 
hand, when introduction of a hole is accompanied by pseudospin-flip of 
other electrons, many states with the same angular momentum are 
connected by the Coulomb interaction, and the charge density of the hole 
has a wider distribution. 
This connected quantum state corresponds to a skyrmion. 
When the charge is confined like a wave function $\phi_{0, X}$ given by Eq. (\ref{eq:singlewf}), 
the formation of a skyrmion reduces the charge locality. 
At high Landau level $N$, however, a wave function $\phi_{N,X}$ is intrinsically broad, 
so the benefit to form skyrmions is lacking. 
In $M$-layered graphene,  the spinor wave function has the localized component $\phi_{0,X}$ at Landau level $N=M$, so it can drive the system to form skyrmions.

\subsection{Crystal structure}
When Zeeman energy for valley pseudospins does not exist, 
a skyrmion splits into two merons (half-skyrmions). 
Four textures of a meron are possible from the two direction at center and the two vorticities. 
	\begin{figure}[h]
	\centering
	\includegraphics[width=6.0cm,clip,angle=-90]{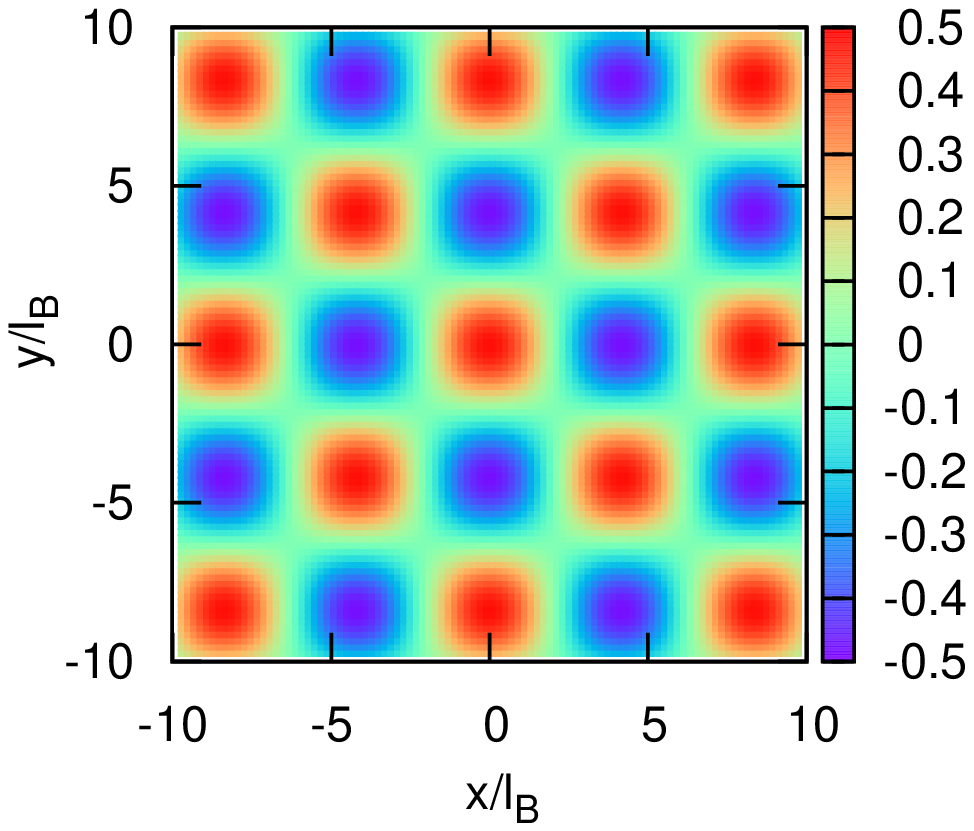}
	\caption{(Color online) $P_z (\bm{r})$ for meron crystal at $\nu_1 = 0.86$ in SLG. }
	\label{fig:MCvpsz}
	\centering
	\includegraphics[width=6.0cm,clip,angle=-90]{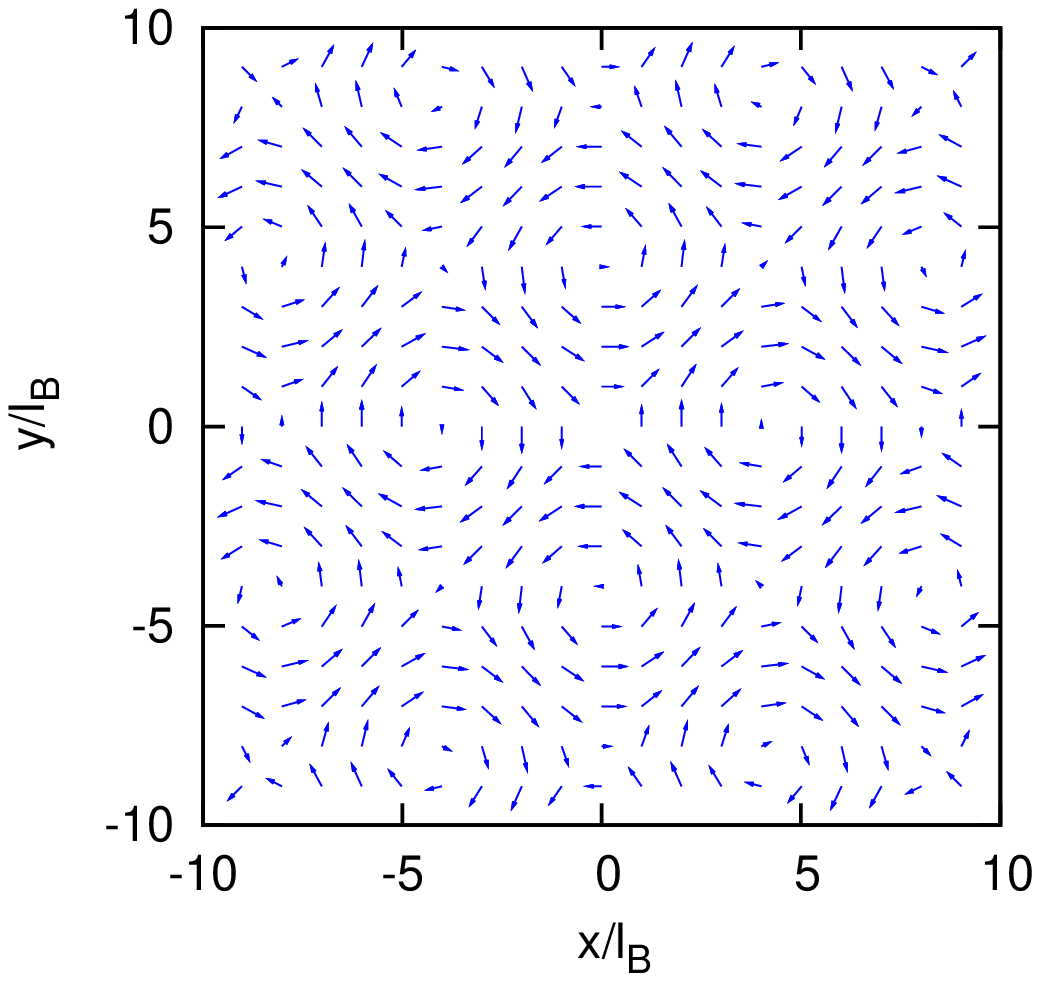}
	\caption{(Color online) The $XY$ orientation of pseudospins for meron crystal at $\nu_1 = 0.86$ in SLG. 
One meron crystal has two types of merons. }
	\label{fig:MCvps2d}
	\end{figure}%
Charge density wave states 
where electrons, holes, or merons form a triangular or square lattice structure are considered. 
The order parameters $\Delta_N^{\sigma, \sigma'} (\bm{Q})$ are defined at points 
\begin{equation}
\bm{Q} = \left[ \left( j + \frac{1}{2}k \right) Q_0, \frac{\sqrt{3}}{2}k Q_0 \right],\ \  \text{for a triangular lattice},
\end{equation} 
\begin{equation}
\bm{Q} =  \left( j Q_0, k Q_0 \right),\ \  \text{for a square lattice}, 
\end{equation}
where $j$ and $k$ are integers. 
The following states are assumed:
\begin{enumerate}
\item{Electron Wigner crystal (eWC) and $n$-electron bubble crystal (eBC$n$): 
a triangular or square lattice with one or $n$ electrons per unit cell. 
The fundamental length in $\bm{q}$ space $Q_0$ is determined from the condition that the CDW has $n$ electrons in a unit cell: $\nu = 2 \pi l_B^2 n/s$. Here, $\nu$ is the  filling factor of electrons and $s$ is the area of a unit cell. 
The low-energy state is pseudospin ferromagnetic because of the Pauli principle. 
}
\item{ Hole Wigner crystal (hWC) and $n$-hole bubble crystal (hBC$n$): 
a triangular or square lattice with one or $n$ holes per unit cell. 
The fundamental length in $\bm{q}$ space $Q_0$ is determined from the condition that the CDW has $n$ holes in a unit cell: $\nu_h = 2 \pi l_B^2 n/s$. Here, $\nu_h$ is the  filling factor of holes and $s$ is the area of a unit cell. 
The low-energy state is pseudospin ferromagnetic because of the Pauli principle. 
}
\item{ Meron crystal (MC): 
a square lattice with four merons of charge $-e/2$ ($\nu < 1$) or $e/2$ ($\nu > 1$) per unit cell, equally spaced. 
A meron pair is equivalent to one skyrmion, so MC can be seen as a state with two skyrmions per unit cell. 
Thus $Q_0$ is determined from the condition that the CDW has $n$ ($=2$) skyrmions in a unit cell: $\nu_h = 2 \pi l_B^2 n/s$. 
The $z$ component of pseudospin density in real space and the vorticity alternate from one site to the next (Figs. \ref{fig:MCvpsz} and \ref{fig:MCvps2d}). 
The $XY$ orientation of pseudospins has $U(1)$ symmetry (Fig.~\ref{fig:MCvps2d}). 
The density distribution in real space is bipartite in layers.
}
\item{ Meron pair crystal (MPC): 
a triangular lattice with four merons per unit cell. 
The merons are not equally spaced and bound into pairs. 
The $Q_0$ is determined from the same way as MC. 
The energy of MPC is similar to MC, 
and has slightly lower energy in the low quasiparticle density regime (close to $\nu_N =1$) in general.    
}
\end{enumerate}

\section{Results\label{sec:Results}}
We use the partial filling factor $\nu_N$ at Landau level $N$; 
thus the total filling factor is given by $\nu = 4N - 2 + \nu_N$ for any-layered graphene. 
Assume that Zeeman splitting is sufficiently large, so the phase diagram for $\nu_N \in [0,2]$ is identical to that for $\nu_N \in [2,4]$. 
Furthermore, the Hamiltonian has electron-hole symmetry around $\nu_N =1$, 
so the phase diagram for $\nu_N > 1$ is caught by alternating  particles for $\nu_N < 1$ to antiparticles. 
Numerical calculation for MC and MPC are done at $ \lvert 1 - \nu_N \rvert \ge 0.06$,  since too many wave vectors are needed to get well-converged solutions at $\nu_N \simeq 1$. 
In the following results, the energies with an accuracy of $10^{-6}$ are presented.
Wigner and bubble crystals are calculated only in triangular symmetry; 
a square lattice generally has higher energy than a triangular one in a low quasiparticle density regime.

It is difficult to get MC solutions close to $\nu_N = 1.0$, 
so we extrapolate the order parameters of the MC solutions in $\nu_N \le 0.94$.
To execute the extrapolation,  
the quantity $\mathcal{F}_{M,N} ( Q ) \Delta^{\sigma, \sigma'}(\bm{Q})$ is fitted by a quadratic curve. 
The energies of the extrapolated MC states are represented in the following figures as a dotted line.  
It is noted that uncertainty remains in the extrapolation especially in the 4-LG case. 
It comes from the relatively low validity of fitting the order parameters which have an inflection point near $\nu_N = 0.92$.

\begin{figure}[h]
	\centering
	\includegraphics[width=5.8cm,clip,angle=-90]{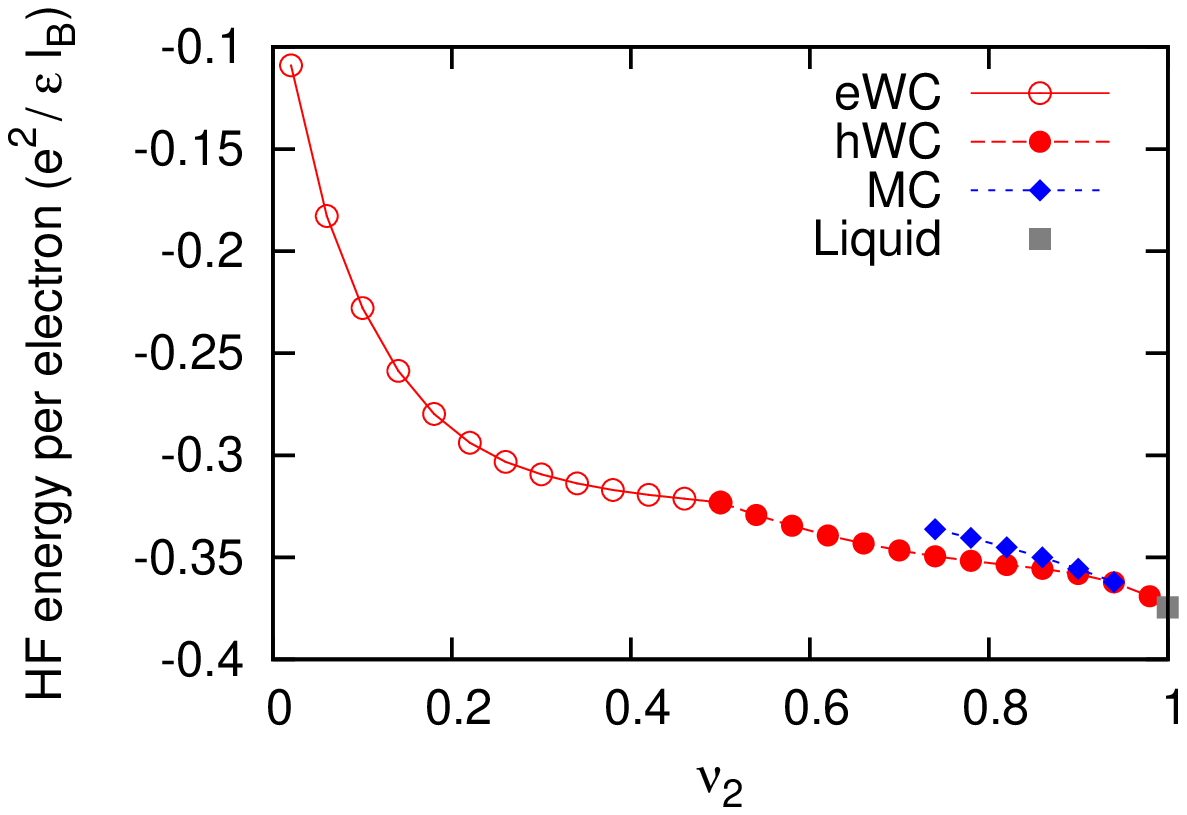}
	\caption{(Color online) Ground-state energy per particle (in units of $e^2/\epsilon l_B$) at Landau level $N=2$ in BLG ($d/l_B = 0$ ). }
	\label{g2l2}
	\centering
	\includegraphics[width=5.8cm,clip,angle=-90]{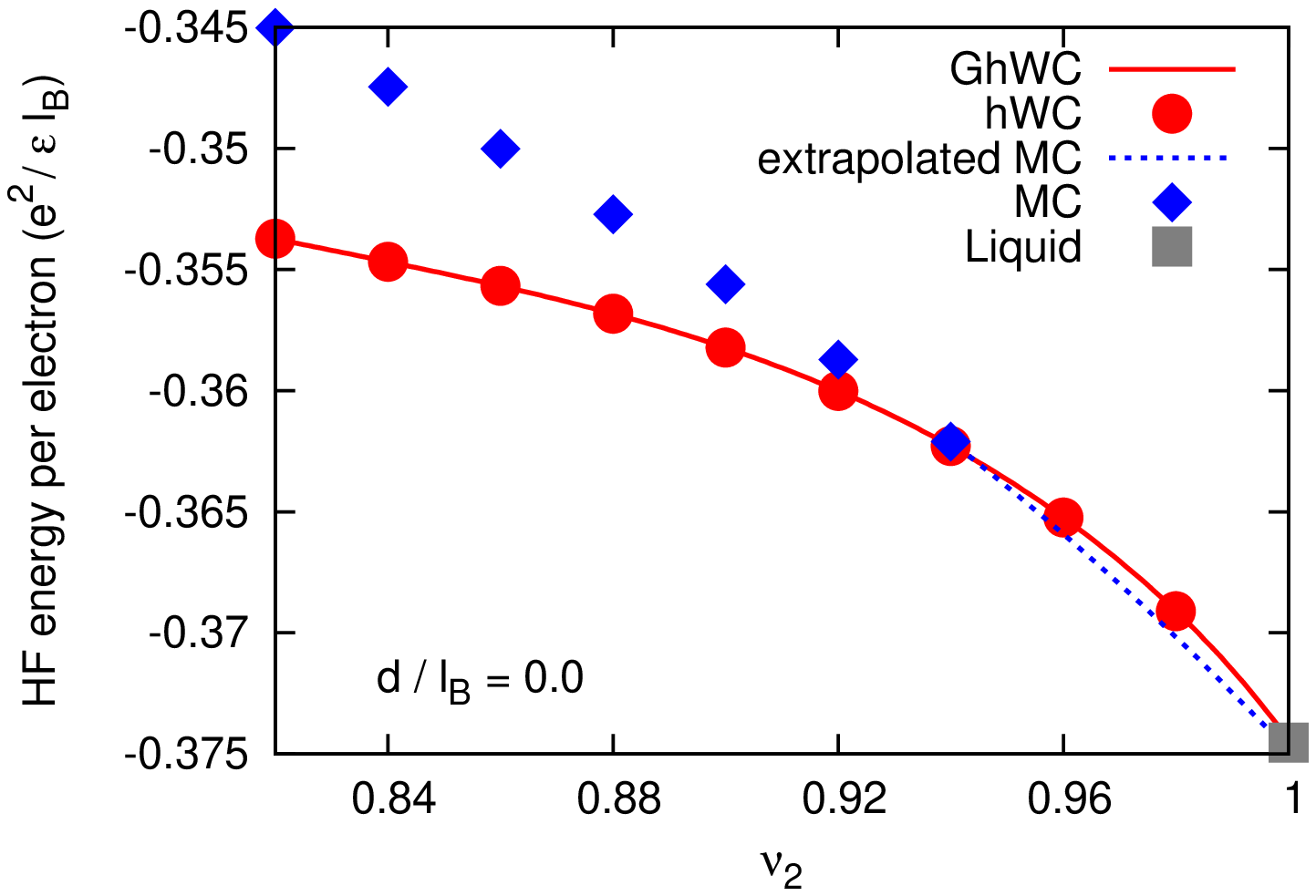}
	\caption{(Color online) Ground-state energy per particle (in units of $e^2/\epsilon l_B$) around filling $\nu_2 = 1$ at Landau level $N=2$ in BLG ($d/l_B = 0$). }
	\label{g2l2high}
\end{figure}

The energies of the valley-concentrated hole Wigner crystal states which are quite accurately approximated by Gaussian form order parameters 
\begin{equation}\begin{split}
& \Delta^{\sigma, \sigma}(0) = \nu_N,  \Delta^{\sigma, \sigma}(\bm{Q}\ne 0) = (\nu_N -1) e^{-Q^2l^2/4}, \\
& \Delta^{\sigma, \bar{\sigma}}(\bm{Q}) = \Delta^{\bar{\sigma}, \sigma}(\bm{Q}) =\Delta^{\bar{\sigma}, \bar{\sigma}}(\bm{Q}) = 0, 
\end{split}\end{equation} 
are represented as ``GhWC'' in the following figures. 
In the vanishing interlayer distance limit, 
the GhWC state with $P_z =(\nu_N^+ - \nu_N^-)/2 = \pm \nu_N/2$ is degenerated to the hWC solutions with $|P_z| < \nu_N /2$ (Figs. \ref{g2l2high}, \ref{g3l3high}, \ref{g4l4high}). 

In the following we show phase diagrams obtained by the present HF
approximation in the whole range of $\nu_N$. It should be remarked that
the true phase diagram should contain regions of incompressible liquid
states that cannot be obtained by the HF approximation. Thus the phase
diagrams are partly incorrect. However, HF calculation gives
qualitatively correct results when the fractional quantum Hall states do
not appear. This is established from comparisons with the results by the
exact diagonalization method~\cite{PhysRevLett.50.1219,
PhysRevB.29.6833, PhysRevLett.100.116802} or the density matrix
renormalization group method.~\cite{PhysRevLett.86.5755} In this paper
we focus on the possibility of meron crystals near $\nu_N=1$, where the
liquid states are not expected, but only the charge-ordered states
compete. Therefore, the following discussion as to the realization of 
the meron crystal is reliable.

\subsection{Single-layer graphene}

The Hartree-Fock (HF) phase diagram of the 2DES in SLG has been obtained and compared with that of the conventional one.~\cite{PhysRevB.75.245414} 
It is shown that Skyrme crystals (MC and MPC) become the ground state around $\nu_N=1$ at Landau levels $N=0$ and $1$.~\cite{PhysRevB.78.085309} 
Considering the excitation energies for SLG (Table~\ref{tableexcitation}), 
it is also possible for the Skyrme crystal phase to occur at $N=2$ and $3$,
but this has not been found yet in a mean-field calculation.
Although the same HF calculation had been done for SLG,~\cite{PhysRevB.75.245414, PhysRevB.78.085309} we executed additional checks and investigated the higher filling regime. 

At Landau level $N=0$, the phase diagram is the following: eWC for $\nu_0 \in [0.10,0.50]$, hWC for $\nu_0 \in [0.50,0.54]$, MC for $\nu_0 \in [0.54,0.63]$, and MPC for $\nu_0 \in [0.63,0.92].$~\cite{PhysRevB.78.085309} 

At Landau level $N=1$, the phase diagram is the following: eWC for $\nu_1 \in [0.10,0.50]$, hWC for $\nu_1 \in [0.50,0.73]$, MC for $\nu_1 \in [0.73,0.84]$, and MPC for $\nu_1 \in [0.84,0.92]$.~\cite{PhysRevB.78.085309} The range of a skyrmionic (MC or MPC) phase is narrower than that of the $N=0$ case. 

At Landau level $N=2$, the phase diagram is the following: eWC for $\nu_2 \in [0.10,0.27]$, eBC2 for $\nu_2 \in  [0.27,0.50]$, hBC2 for $ \nu_2 \in [0.50,0.73]$, and hWC for $ \nu_2 \in [0.73,0.94]$. It is characteristic that 2-electron  (hole) bubble crystals exist around $\nu_2 = 0.50$. The skyrmionic ground state is not seen in the range $\nu_2 \le 0.94$. 
The extrapolating analysis, however, suggests that the hWC and MC state are almost degenerated in $\nu_2 \in [0.94 ,1.0]$.

At Landau level $N=3$, the phase diagram is the following: eWC for $\nu_3 \in [0.10,0.20]$, eBC2 for $\nu_3 \in  [0.20,0.30]$, eBC3 for $\nu_3 \in  [0.30,0.50]$, 
hBC3 for $\nu_3 \in  [0.50,0.70]$, hBC2 for $ \nu_3 \in [0.70,0.80]$, and 
hWC for $ \nu_3 \in [0.80,0.94]$. The skyrmionic state does not appear in $\nu \le 0.94$. 

Although the MC and MPC solutions are not found in $\nu_N \le 0.94$ for SLG at $N=2$ and $3$, 
the Skyrme crystal is expected to have lower energy in the immediate vicinity of $\nu_N = 1$ from the analysis in Sec. \ref{subsec:excitation}.

\subsection{Landau level $N=2$ in bilayer graphene}
The HF calculation for degenerated zero-energy Landau levels $N=0, 1$ in BLG 
suggests that the Skyrme crystal states of real spin or orbital pseudospin occur.~\cite{PhysRevB.82.245307} 

In what follows, the results for the first excited Landau level $N =2$ in BLG are presented. 
Figure~\ref{g2l2} shows the energies per electron for several crystal structures. 
It shows the following sequence of ground states: 
eWC for $\nu_2 \in [0.10,0.50]$,  hWC for $ \nu_2 \in [0.50, 0.94]$. 
The bubble state does not appear unlike the phase diagram at Landau level $N=2$ for SLG. 
Figure~\ref{g2l2high} shows the energies in the area close to $\nu_2=1$. 
The extrapolated energy of the MC solutions have lower value than hWC for $ \nu_2 \in [0.94, 1.0]$.

\begin{figure}
	\centering
	\includegraphics[width=5.8cm,clip,angle=-90]{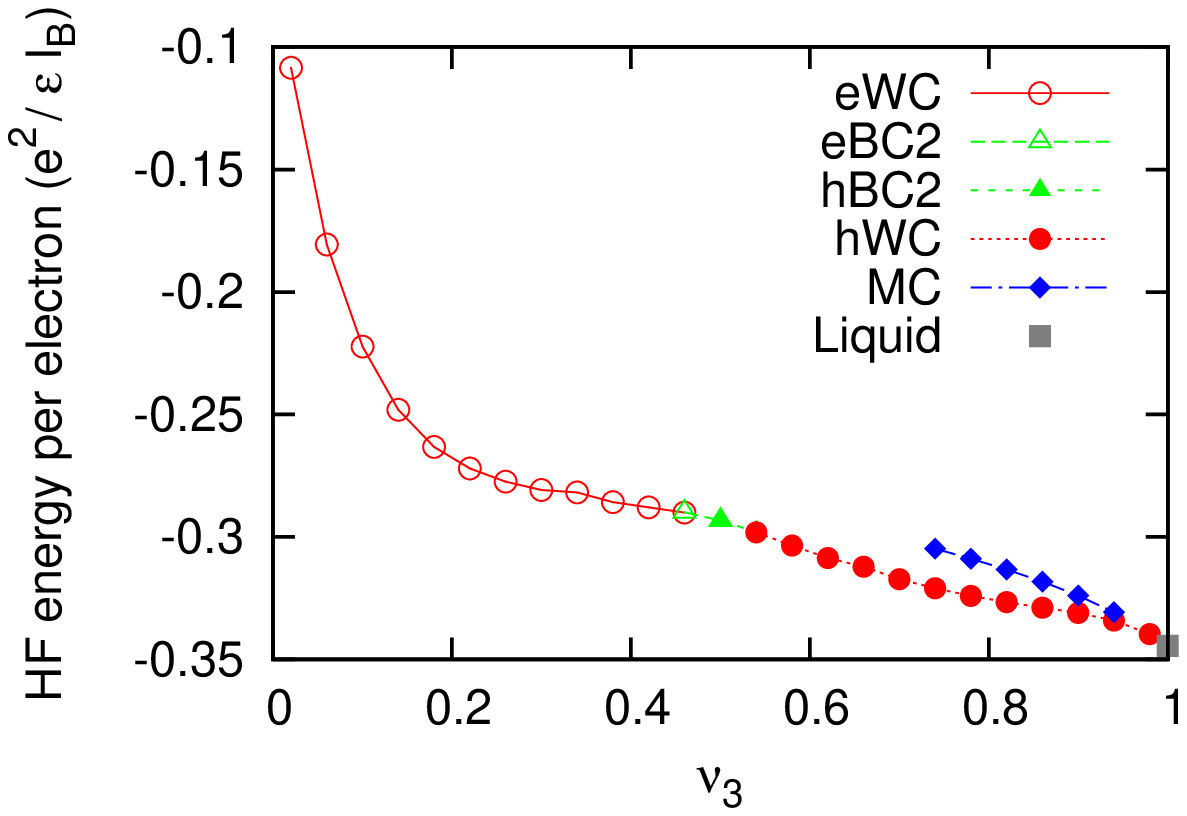}
	\caption{(Color online) Ground-state energy per particle (in units of $e^2/\epsilon l_B$) at Landau level $N=3$ in tri-LG ($d/l_B = 0$ ). }
	\label{g3l3}
	\centering
	\includegraphics[width=5.8cm,clip,angle=-90]{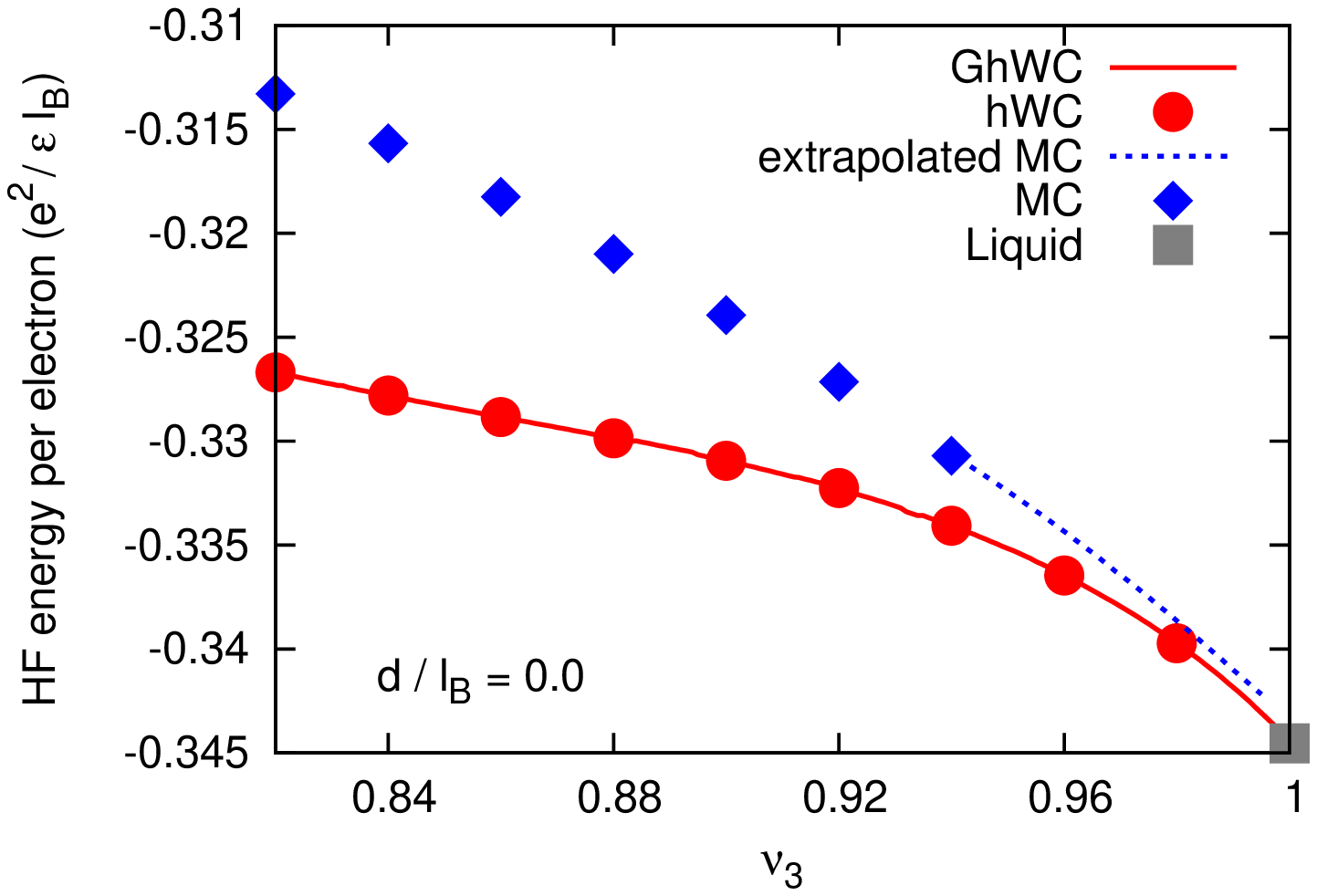}
	\caption{(Color online) Ground-state energy per particle (in units of $e^2/\epsilon l_B$) around filling $\nu_3 = 1$ at Landau level $N=3$ in tri-LG ($d/l_B = 0$). }
	\label{g3l3high}
\end{figure}%

\subsection{Landau level $N=3$ in trilayer graphene}

Figure~\ref{g3l3} shows the energies per electron for several crystal structures at Landau level $N=3$ in tri-LG. 
It shows the following sequence of ground states: 
eWC for $\nu_3 \in [0.10,0.46]$, eBC2 for $\nu_3 \in  [0.46,0.50]$, 
hBC2 for $ \nu_3 \in [0.50,0.54]$, hWC for $ \nu_3 \in [0.54,0.94]$.
Although the bubble states are found around $\nu_3 = 0.50$, 
its range is narrower than that of the $N=3$ case in SLG. 
The skyrmionic ground state is not seen in the range $\nu \le 0.94$. 
Figure~\ref{g3l3high} shows the energies near $\nu_3=1$. 
The extrapolated states of the MC solutions have an energy close to that of  hWCs in the vicinity of $ \nu_3=1$.

\subsection{Landau level $N=4$ for tetralayer graphene}

\begin{figure}[t]
	\centering
	\includegraphics[width=5.8cm,clip,angle=-90]{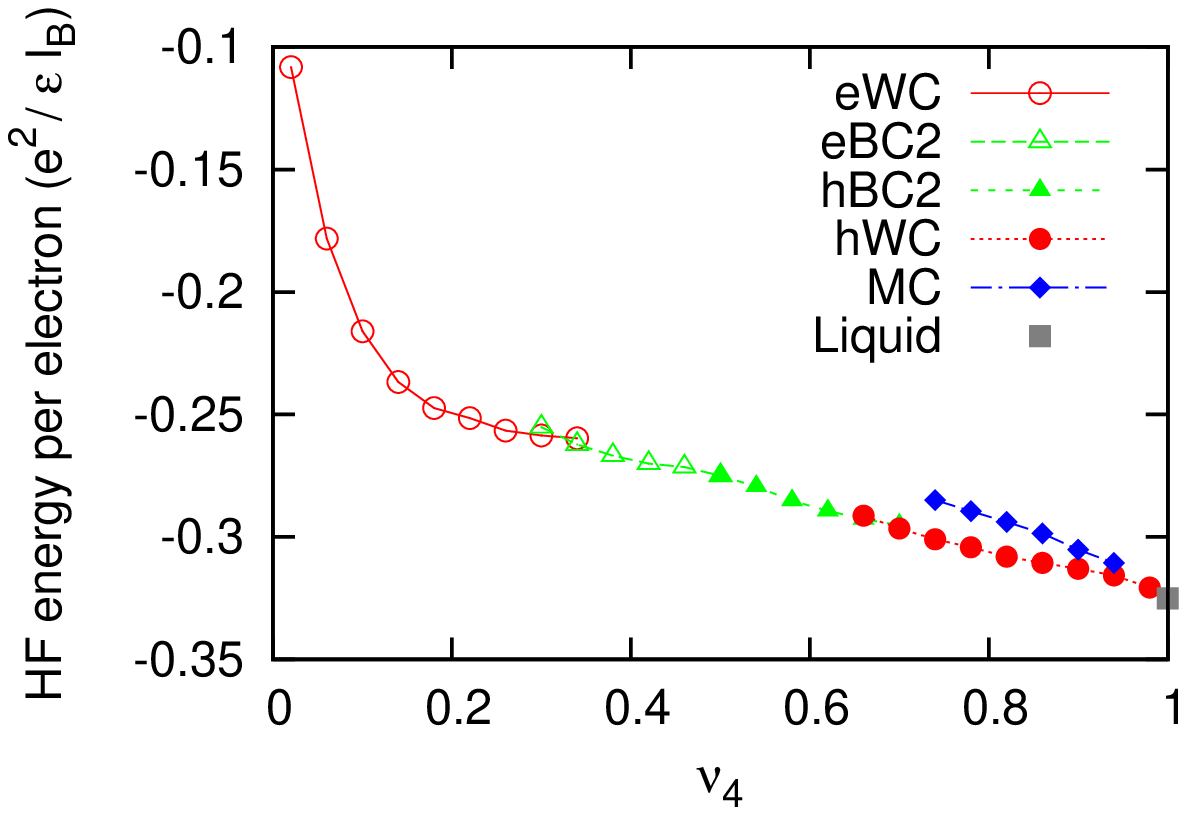}
	\caption{(Color online) Ground-state energy per particle (in units of $e^2/\epsilon l_B$) at Landau level $N=4$ in tetra-LG ($d/l_B = 0$ ). }
	\label{g4l4}
	\includegraphics[width=5.8cm,clip,angle=-90]{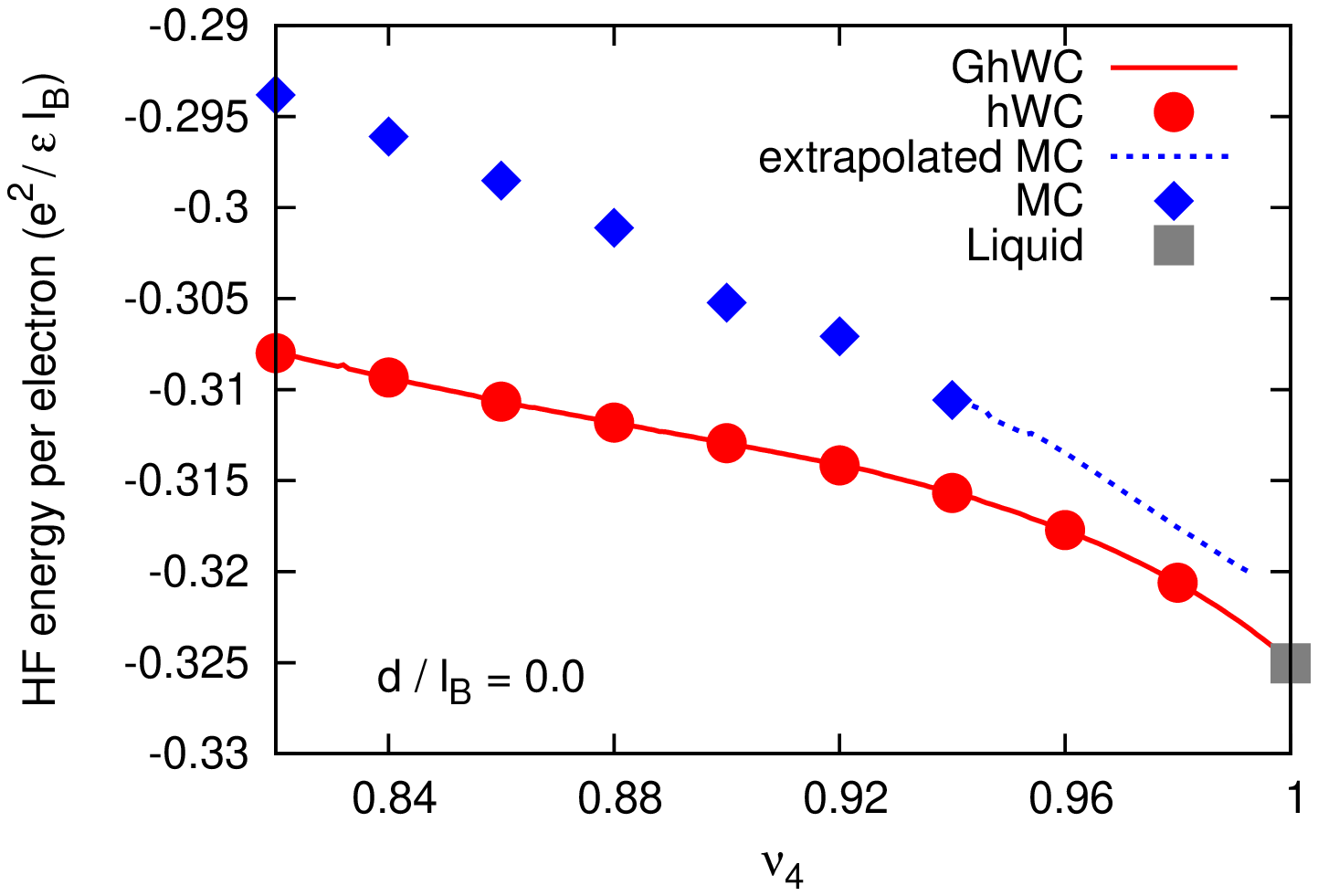}
	\caption{(Color online) Ground-state energy per particle (in units of $e^2/\epsilon l_B$) around filling $\nu_4 = 1$ at Landau level $N=4$ in tetra-LG ($d/l_B = 0$ ). }
	\label{g4l4high}
\end{figure}%
Figure~\ref{g4l4} shows the energies per electron for various crystal structures at Landau level $N=4$ for tetra-LG. 
It shows the following sequence of ground states: 
eWC for $\nu_4 \in [0.10,0.32]$, eBC2 for $\nu_4 \in  [0.32,0.50]$, 
hBC2 for $\nu_4 \in [0.50,0.68]$, and hWC for $ \nu_4 \in [0.68,0.94]$.
The bubble states are found in the broad range $\nu_4 \in [0.32,0.68]$. 
The skyrmionic ground state is not seen in the range $\nu_4 \le 0.94$. 
Figure~\ref{g4l4high} shows the energies near $\nu_4=1$. 
The extrapolated states of the MC solutions have higher energy than that of  hWCs in the vicinity of $ \nu_4=1$.  
As previously mentioned, however, the extrapolation method is no longer valid in tetra-LG. 
According to the analysis in Sec. \ref{subsec:excitation}, 
the Skyrme crystal is expected to have the lowest energy in the vicinity of $\nu_4 = 1$.

\section{Discussion\label{sec:Discussion}}
The Hartree-Fock (HF) calculation suggests that 
the meron crystal (MC) phase appears at Landau level $N=2$ in BLG, 
and hole Wigner crystal (hWC) and MC states are degenerated around $\nu_N =1$ at Landau level $N=2$ in SLG, $N=3$ in tri-LG in the case of vanishing interlayer distance. 
Our calculation strongly suggests that meron pair crystals (MPCs) will have lower energy in the immediate vicinity of $\nu_N=1$ for its triangular symmetry. 

Skyrme crystals (MC and MPC) have a charge distribution and collective mode different from that of hWC, so these states can be distinguished by transport properties~\cite{PhysRevLett.65.2189, PhysRevLett.82.394} and a microwave absorption spectrum.~\cite{PhysRevB.78.085309, PhysRevLett.104.226801} 
Furthermore, the CDWs exist in outer layers, 
so the local density of states (LDOS) can be measured in a spectroscopic manner.~\cite{PhysRevB.80.195414} 
In this paper, the unidirectional stripe phase is not considered. 
The stripe phase, however, also will appear around $\nu_N =0.5$ in BLG and $M$-LG, as SLG~\cite{PhysRevB.75.245414} and conventional 2DES if fractional quantum Hall states are not realized. 
Such a state, if exists, will be identified by anisotropic conduction.~\cite{PhysRevLett.82.394} 

Although we ignored the effects of disorder, a finite valley Zeeman energy, and Landau-level transitions, it is unclear how these affect degeneracy of hWC and MC around $\nu_N =1$.   
In particular, Landau-level transitions in BLG and $M$-LG are larger than that of SLG under a magnetic field $B \sim 10$ T. 
The gap near a charge neutrality point is 
$\Delta_1 = \sqrt{2}\hbar v_F / l_B \sim 380\sqrt{B/\mathrm{T} }\  \mathrm{K}$ for SLG,
$\Delta_2 = \sqrt{2}\hbar \omega_c \sim 45 \times (B /\mathrm{T})\  \mathrm{K}$ for BLG, 
$\Delta_3 = \sqrt{6} \hbar \omega_3 \sim 6.6 \times (B /\mathrm{T})^{3/2}\  \mathrm{K}$ for tri-LG, and 
$\Delta_4 = \sqrt{24} \hbar \omega_4 \sim 1.1 \times (B /\mathrm{T})^2 \ \mathrm{K}$ for tetra-LG. 
The typical Coulomb energy is 
$E_{\textit{C}}=e^2/ \epsilon l_B \sim 100 \sqrt{B/\mathrm{T}} \ \mathrm{K}$. 
For SLG, the ratio $E_{\textit{C}} / \Delta_1 = 0.39$ is independent of field $B$. 
In this case it is shown that the Landau-level mixing does not change the CDW phase diagram (except the skyrmionic crystal).~\cite{PhysRevB.77.205426} 
For BLG and $M$-LG, high magnetic fields are needed to achieve a comparable ratio: 
$B \sim 70\  \mathrm{T}$ for BLG, $60\ \mathrm{T}$ for tri-LG, and $50\ \mathrm{T}$ for tetra-LG.

It has been pointed out that anisotropy in the pseudospin arises in the
order of $a/l_B$.~\cite{PhysRevB.74.161407, PhysRevB.74.075422, PhysRevLett.98.196806, PhysRevLett.98.016803, RevModPhys.83.1193} 
In the multilayer graphene, finite layer separation also brings anisotropy. 
This anisotropy is quite small, since $a < d \ll l_B$, 
but may have some effect when the energies of two phases are quite close. 
We have done a calculation taking into account only the effect of $d/l_B$ as a preliminary investigation, 
and found that the finite $d$ is slightly unfavorable for the Skyrme crystal. 
However, to obtain a definite conclusion for the effect of finite $d$, 
we need to take into account the effect of $a$ also. Such calculation is left for future investigation. 

\begin{acknowledgments}
Y.S. thanks R. C\^ot\'e for helping him to find self-consistent
solutions to MC states. 
The numerical calculation was done by SR11000 at Information Technology Center,  University of Tokyo. 
\end{acknowledgments}


\begin{thebibliography}{99}
\bibitem{note1}
Although ABC-stacked graphene has a different stacking sequence from common graphite,  
we studied this possible sequence here.
\end{thebibliography}
\end{document}